\documentclass[prd,aps, tightenlines, preprint, preprintnumbers, showpacs, nofootinbib,superscriptaddress]{revtex4}

\usepackage{graphicx}
\usepackage{amsmath,amsthm,amssymb}
\usepackage{mathtools}
\usepackage[usenames,dvipsnames]{color}

\begin{document}
\title{ Sudakov suppression of  the Balitsky-Kovchegov kernel }

\author{Du-xin Zheng}
\affiliation{\normalsize\it  Key Laboratory of Particle Physics and
Particle Irradiation (MOE), Institute of frontier and interdisciplinary science, Shandong
University,  QingDao, Shandong 266237, China}
\author{Jian Zhou}
\affiliation{\normalsize\it  Key Laboratory of Particle Physics and
Particle Irradiation (MOE), Institute of frontier and interdisciplinary science, Shandong
University,  QingDao, Shandong 266237, China}

\begin{abstract}
 To sum high-energy leading logarithms in a consistent way,
 one has to impose the strong ordering in both projectile rapidity and dense target rapidity
simultaneously, which results in a kinematically improved  Balitsky-Kovchegov(BK) equation.
 We find that beyond this strong ordering region,
 the important sub-leading double  logarithms arise at high order due to the incomplete
 cancellation  between real corrections and virtual corrections in a t-channel calculation.
 Based on this observation,
 we further argue that these double logarithms are the Sudakov type ones, and thus can be
 resummed into an exponential leading to a Sudakov suppressed BK equation.
\end{abstract}

\pacs{...} \maketitle

\section{introduction}
The non-linear evolution equation---the Balitsky-JIMWLK equation~\cite{Balitsky:1995ub,JalilianMarian:1997jx}
 or its mean-field truncation: the BK equation~\cite{Balitsky:1995ub,Kovchegov:1999yj}
 plays a central role in studying saturation physics. The leading order evolution kernels have been derived
 about two decades ago.
 To make realistic predictions for observables in various high energy scattering processes
 at RHIC, LHC and the future EIC,
 it is necessary to utilize the NLO version of the BK equation in phenomenology studies.
  Though the  calculation of NLO BK kernel is extremely complicated,
  it has been eventually achieved in Ref.~\cite{Balitsky:2008zza}.
 Later, there was also  attempt to extend the NLO analysis to
 the Balitsky-JIMWLK equation~\cite{Balitsky:2013fea,Kovner:2013ona}.

 However, when trying to solve the NLO BK equation numerically~\cite{Lappi:2015fma},
 it was found to be unstable.  This is essentially due to the
  NLO correction enhanced by double logarithms, which becomes very large when the parent  dipole is much
 smaller than the daughter dipole. As a matter of fact,
 the similar problem was earlier observed in the context  of the
 BFKL dynamics~\cite{Kuraev:1977fs,Balitsky:1978ic} as well.
 To cure the instability issue and improve the convergence of the perturbation series,
 significant efforts were  then devoted to resum these large double logarithms to all orders
  in the dilute limit~\cite{Salam:1998tj,Ciafaloni:1999yw,Ciafaloni:2003rd,Altarelli:2005ni,Ciafaloni:2007gf}
   and in the saturation case~\cite{Beuf:2014uia,Iancu:2015vea,Iancu:2015joa,Lappi:2016fmu,Ducloue:2019ezk}.
   The common feature of various proposed resummation schemes
  is to implement a kinematical constraint for successive  gluon radiations during small $x$ evolution.
It has been argued that these large double logarithms can be effectively resummed to all orders
 via introducing a kinematically improved BK equation.

   It is indeed a natural idea to impose a kinematical constraint in the evolution kernels
   as the derivations of the BK/BFKL equations rely on making systematical kinematical approximations
    in strong rapidity ordering region. In particular, to construct a self-consistent small $x$
    evolution, the strong ordering in both projectile rapidity and dense target
    rapidity must be satisfied simultaneously. The BK equation is conventionally  formulated as an
    evolution with the projectile rapidity. The strong ordering in projectile rapidity is
  realized by construction in this case, whereas the strong target rapidity ordering is not
  automatically guaranteed. One way of enforcing the target rapidity ordering is to insert a theta
  function in the LO BK kernel.   The most important part of the
   double logarithms arises at  high order corrections is shown to be  recovered from a such non-local BK equation~\cite{Beuf:2014uia}.
   Alternatively, the double-logarithmic corrections can be explicitly resummed in the kernel and give rise
   to a collinearly-improved BK equation~\cite{Iancu:2015vea,Iancu:2015joa}. These two methods are equivalent up to the leading double logarithm
   accuracy. More recently,  a BK equation with target rapidity being the
   evolution variable  was argued to be superior than that describing projectile rapidity evolution, in the
   sense that the resummed version of the former one has the less scheme dependence~\cite{Ducloue:2019ezk}.

 In this paper, we focus on investigating the contributions from the phase space region
 beyond the aforementioned strong ordering region.  We adopt
  target rapidity evolution point of view in our analysis. In this scenario,
 the so-called anti-collinear logarithm
  appears in the full two loop BK kernel~\cite{Balitsky:2008zza} is absent at the NLO.
 Instead, a  double logarithm generated from the phase space region of interest
 becomes a large correction when one of the daughter dipoles
is much smaller than the parent dipole. This finding is consistent with
 what the authors of the paper~\cite{Ducloue:2019ezk} observed.
 Moreover, the fact that the double logarithm terms result from the
 incomplete cancellation between real corrections and virtual corrections leads us to identify it
 as the Sudakov logarithm. Consequently, these logarithms
  can be resummed into an exponential following the standard
 procedure, and give rise to a Sudakov suppressed BK equation.
 However, quite puzzlingly,
 the expansion of the Sudakov suppressed non-local BK equation to the first non-trivial order
  would produce a double logarithm term that is different from the corresponding one
  appears in the NLO BK equation with the target rapidity being the evolution variable~\cite{Ducloue:2019ezk}.
 Though this mismatch must be clarified at some point, we are not able to offer a solution for the moment, but
  aim at investigating it thoroughly in a future work.

  In  recent years, the study of Sudakov resummation in the context of small $x$ physics is actually becoming
  a topical issue. This was first initiated by the authors of Refs.~\cite{Mueller:2012uf}, in which a joint resummation
  formalism  was developed to compute physical observables in high energy scattering processes
   involving at least three well separated scales.
   In order to formulate the joint $k_t$ and small $x$ resummation in
   the conventional TMD factorization framework, the Collins-Soper evolution and the
   scale dependence of gluon TMD in the small $x$ limit
   was later investigated in Refs.~\cite{Zhou:2016tfe,Xiao:2017yya,Zhou:2018lfq}.
 Such an effective TMD factorization in the small $x$ limit  has also been applied
 in phenomenology studies~\cite{Zheng:2014vka,vanHameren:2014ala,vanHameren:2019ysa,Boer:2017xpy,Dong:2018wsp}.
   We notice that the other formulations
   exist in the literatures~\cite{Balitsky:2015qba,Balitsky:2019ayf,Marzani:2015oyb}.

   The papers is organized as follows. In the next section, we first re-derive the LO BK equation from a
   t-channel calculation. Following a short discussion about kinematically improved BK equation, we identify
   the phase space region where the large double logarithm can be produced.
 We then proceed to analyze the  dynamical origin  of the double logarithm, and argue how to systemically
 resum them to all orders. The paper is summarized in Sec.III.

\section{derivation of the double logarithm correction to the BK kernel at two loop order and beyond}
The large logarithm correction to the BK kernel arises at two loop order.
In order to fix the notations and set up the baseline for the two
loop calculation, we first briefly review the derivation  of the
leading order BK evolution kernel. To this end, we compute the NLO
correction to the the dipole matrix element defined below,
 \begin{eqnarray}
\frac{1}{N_c} \langle {\rm Tr} [ U(x_\perp)U^\dag(y_\perp) ] \rangle
\end{eqnarray}
The Wilson line is defined as,
 \begin{eqnarray}
U(x_\perp)= {\cal P} \exp \left \{ ig_s \int_{-\infty}^{+\infty}
dx^- A^+(x^-,x_\perp)T^c \right \}
\end{eqnarray}
where the plus($+$) and the minus($-$) symbols represent   the commonly defined light
cone components of four momenta.
At the leading order, the dipole amplitude $\langle U(x_\perp)
U^\dag(y_\perp) \rangle $ is independent of rapidity. It is
normally determined in the McLerran-Venugopalan model~\cite{McLerran:1993ni} as the initial
condition. The large logarithm $\ln \frac{1}{x_g}$  shows up in the
one loop correction to the above matrix element. These logarithm
terms at high orders can be absorbed into the dipole amplitude and give rise to the
energy dependence of the correlator.

To avoid the interaction between the radiated gluon and color source inside target,
our calculation is performed in the light cone gauge ($A^-=0$), in which gluon propagator reads,
\begin{eqnarray}
\left ( -g^{\mu \nu} +\frac{P^\mu k^\nu+P^\nu k^\mu}{k \cdot
P-i\epsilon}\right ) \frac{i}{k^2+i\epsilon}
\end{eqnarray}
where the prescription $\frac{1}{k \cdot P-i\epsilon}$ for regulating the light cone divergence
is proven to be the most convenient choice for our calculation.

\begin{figure}[htpb]
\includegraphics[angle=0,scale=0.60]{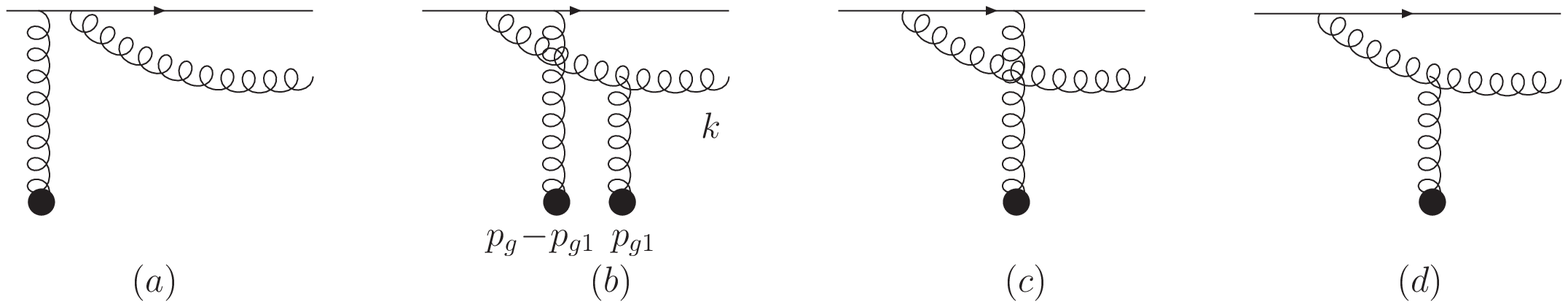}
\includegraphics[angle=0,scale=0.7]{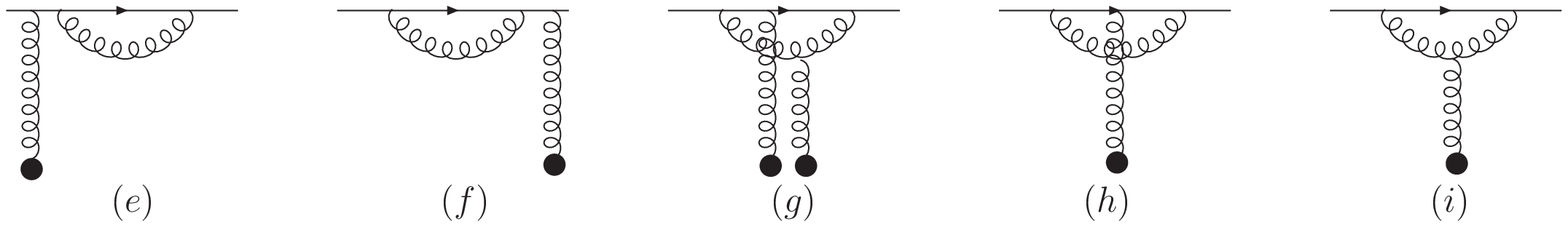}
\caption{ Real¡¡ and virtual diagrams contributing to the BK kernel at the leading
order. The black dot stands for multiple gluon re-scattering . The
straight solid line represents the Wilson line in the fundamental
representation.
 } \label{fig1}
\end{figure}

The real contribution to the BK kernel at the leading order comes from graphs shown in Figs.[\ref{fig1}a-d].
 In momentum space, the amplitude of Fig.[\ref{fig1}a] is given by,
\begin{eqnarray}
{\cal M}_{1a} = -ig_s\frac{2 k_\perp \cdot \epsilon^*_\perp}{k_\perp^2}
 t^a \left [ U(p_{g\perp})-(2\pi)^2 \delta^2(p_{g\perp}) \right ]
\end{eqnarray}
The amplitude of diagrams Fig.[\ref{fig1}b], Fig.[\ref{fig1}c] and Fig.[\ref{fig1}d] read,
\begin{eqnarray}
{\cal M}_{1b} +{\cal M}_{1c}+{\cal M}_{1d}&=& -ig_s
 \int \frac{d^2 p_{g1\perp}}{(2\pi)^2} \frac{2(k_\perp-p_{g1\perp})\cdot \epsilon^*_\perp}
 {(k_\perp-p_{g1\perp})^2}
 \nonumber \\&&\times
 \left [ U(p_{g\perp}-p_{g1\perp})t^b\tilde U_{ab}(p_{g1\perp})
 -(2\pi)^4 t^a\delta^2(p_{g1\perp})\delta^2(p_{g\perp}-p_{g1\perp}) \right ]
\end{eqnarray}
where $\tilde U(p_{g1\perp})$ is the Fourier transform of the path ordered Wilson lines in the adjoint representation.
After taking the Fourier transform, the squared summation of the amplitudes Figs.[\ref{fig1}a-\ref{fig1}d] in the coordinator space
is expressed as,
\begin{eqnarray}
\frac{g_s^2}{2\pi^2}\frac{(x_\perp\!-z_\perp)\! \cdot \! (y_\perp\!-z_\perp) }{(x_\perp\!-z_\perp)^2(y_\perp\!-z_\perp)^2}
\left \{ 2{\rm Tr}\left [  U^\dag(y_\perp) t^a t^a U(x_\perp) \right ]\!
- 2{\rm Tr}\left [  U^\dag(y_\perp)U(z_\perp) t^a U^\dag(z_\perp) U(x_\perp)t^a  \right ]
\right \}
\end{eqnarray}
To arrive at the above expression, we have used the formula
  $U^\dag(z_\perp) t^a U(z_\perp)=t^b \tilde U_{ab}(z_\perp)$.
By further employing the Fierz identity,
\begin{eqnarray}
&&{\rm Tr}\left [  U^\dag(y_\perp)U(z_\perp) t^a U^\dag(z_\perp) U(x_\perp)t^a  \right ]
\nonumber \\ &&= \frac{1}{2} {\rm Tr}\left [  U^\dag(y_\perp)U(z_\perp)] \right ]
{\rm Tr}\left [ U^\dag(z_\perp) U(x_\perp)  \right ]-
\frac{1}{2N_c}{\rm Tr}\left [  U^\dag(y_\perp) U(x_\perp)  \right ]
\end{eqnarray}
the real contribution to the BK kernel can be cast into a familiar form,
\begin{eqnarray}
\frac{g_s^2}{2\pi^2}\frac{(x_\perp\!-z_\perp)\! \cdot \! (y_\perp\!-z_\perp) }{(x_\perp\!-z_\perp)^2(y_\perp\!-z_\perp)^2}
\left \{ N_c {\rm Tr}\left [  U^\dag(y_\perp)  U(x_\perp) \right ]\!
- \!{\rm Tr}\left [  U^\dag(y_\perp)U(z_\perp) \right] {\rm Tr}\left [ U^\dag(z_\perp) U(x_\perp) \right ]
\right \}
\end{eqnarray}

Similarly, virtual graphs illustrated in Figs.[\ref{fig1}e-\ref{fig1}i]  can be readily computed.
 Combining the real contribution and the virtual contribution and carrying out phase space integration,
 one ends up with,
\begin{eqnarray}
 \langle {\rm Tr} [ U(x_\perp)U^\dag(y_\perp) ]\rangle &=&
\! \langle {\rm Tr} [ U(x_\perp)U^\dag(y_\perp) ]\rangle|_{LO}+
\frac{\alpha_s}{2\pi^2} \!
\int \! \frac{d k^+}{k^+} \int \!\! d^2 z_\perp \frac{(x_\perp-y_\perp)^2}{(x_\perp-z_\perp)^2(y_\perp-z_\perp)^2}
\nonumber \\&\times& \!
\left \{
 {\rm Tr}\left [  U^\dag(y_\perp)U(z_\perp) \right] {\rm Tr}\left [ U^\dag(z_\perp) U(x_\perp) \right ]
- N_c {\rm Tr}\left [  U^\dag(y_\perp)  U(x_\perp) \right ]\right \}
\end{eqnarray}
which matches the first step iteration of the LO BK equation. There are various ways to regulate
 the light cone divergence for $k^+$ integration in the above equation. As long as the calculation
 is performed in the leading high energy logarithm  accuracy, all regularization schemes are equivalent.
 However, they start to deviate from each other by some sub-leading logarithms at high order.
 For the  current purpose, we do not need to specify the regularization scheme as the double logarithm
 under investigation has different dynamical origin.

\subsection{Non-local BK equation and kinematical constraints }

\begin{figure}[htpb]
\includegraphics[angle=0,scale=0.8]{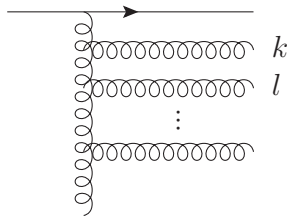}
\caption{ Strong rapidity ordering in successive gluon radiations.
 } \label{cascade}
\end{figure}

When deriving a small $x$ evolution equations like BFKL, BK or B-JIMWLK,
 it is crucial to impose a strong  minus momenta ordering $\bar P^- \gg k^-\gg l^-...$
 from a  projectile evolution point of view  as shown in Fig.\ref{cascade},
 while a strong plus momenta ordering  $ k^+ \ll l^+ ... \ll P^+$ is required
 if the problem is formulated as the evolution of a dense target.
 These two kinematical constraints are compatible with each other so long as
 the transverse momenta carried by the successive radiated gluons are of
 the same order $k_\perp^2 \sim l_\perp^2$. However, this condition can not be always met
 because transverse momentum is integrated over the whole phase space region in the kernel of the high-energy
evolution equations. Instead, it has recently been  recognized~\cite{Beuf:2014uia} that it is necessary
to  have the strong plus momenta ordering and minus momenta ordering simultaneously in order to
generate large high energy logarithm from each gluon emission.

In the present paper, we study the NLO BK evolution in $\ln \frac{1}{k^+}$, namely the
evolution of dense target, for which case the plus momenta are properly ordered by construction.
 One then should add a theta function in the kernel of the BFKL(or BK) equation in order to
 impose the $k^-$ ordering not already guarantied by the choice of evolution variable.
 The various ways of enforcing  such kinematical constraint have been proposed in the literatures~\cite{Kutak:2011fu,Beuf:2014uia,Iancu:2015vea,Iancu:2015joa,Lappi:2016fmu,Ducloue:2019ezk}.
 These kinematically improved
 evolution equations effectively resum  some large high order corrections  to the BFKL/BK kernel to all orders.
 And in general, a kinematically improved BK equation  leads to a more stable small $x$ evolution
  in comparison with the use of the NLO BK kernel without resumming large transverse logarithms.

However, there exists a phase space region  satisfying both conditions
 $ k^+ \gg l^+$ and $ k^-\gg l^-$ in which QCD dynamics is not yet explored.
 If transverse momenta carried by  the successive emitted gluons are of the same order,
 this phase space region basically shrinks to zero. When $k_\perp^2 \gg l_\perp^2$,
 the corresponding phase  region is sufficiently large to potentially induce large logarithm contributions at NLO.
 This can be best seen from the following phase space integration,
\begin{eqnarray}
\int \frac{d^2l_\perp}{l_\perp^2}
\int^{k^+ }_{\frac{l_\perp^2}{k_\perp^2}k^+} \frac{dl^+}{l^+}=\int \frac{d^2l_\perp}{l_\perp^2}
\ln \frac{k_\perp^2}{l_\perp^2}
\end{eqnarray}
where the lower integration limit for $l^+$ integration is determined
 according to the constraint $k^- \gg l^-$.
The denominators $\frac{1}{l^+}\frac{1}{l_\perp^2}$ in the above formula
 represent a typical structure arises from a NLO real correction with a soft gluon emitting from an external line.
The infrared divergence from the real correction will be canceled out when combining with virtual corrections as explained later.
But such cancellation is not complete in the sense that we are left with a double logarithm term.
 Therefore, on top of the non-local BK equation which already takes care part of large logarithm contributions,
 one still need to resum these Sudakov type logarithm terms.
  And once these large logarithms are identified as the typical Sudakov ones,
  they can be resummed into an exponential by the standard method.

\begin{figure}[htpb]
\includegraphics[angle=0,scale=0.6]{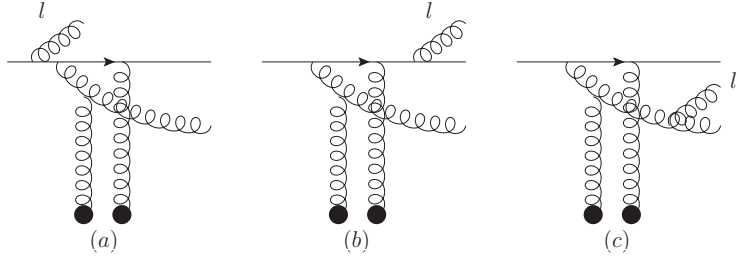}
\caption{ Soft gluon($l$) radiation diagrams.
Diagrams with the soft gluon radiated from the incoming gluon leg do not produce large
Sudakov logarithm.} \label{amplitude}
\end{figure}

In the following analysis, we focus on the NLO contributions from the phase space region
$ k^+ \gg l^+$ and $ k^-\gg l^-$ as the contributions from kinematical regions outside
this slice of phase space has been summarized into non-local BK equation.  We start presenting
 the detailed NLO calculations by writing down the amplitudes of graphs illustrated in Fig.\ref{amplitude}.
 Since all four components of $l$ are much smaller than any scales in the problem under consideration,
 the soft gluon only can be emitted from external lines. The insertion of the soft gluon to an internal
 propagator would lead to a power suppressed contribution. Moreover, it is well known that
 in a light cone gauge calculation, the diagrams with a soft gluon emitting from the incoming gluon lines
 do not produce any Sudakov type logarithm. Some sample soft gluon radiation
 diagrams which could potentially generate the large double logarithm are shown in Fig.\ref{amplitude}.
 Applying  the soft gluon approximation, the dominant contributions of the real correction are given by
\begin{eqnarray}
{\rm Fig.3a} &\propto &-ig_s \frac{2 n^\mu}{2l \cdot n } e^{-ix_\perp \cdot l_\perp}{\cal M}_{LO}^c t^a
\\
{\rm Fig.3b} &\propto& ig_s \frac{2n^\mu}{2l \cdot n} e^{-ix_\perp \cdot l_\perp}t^a{\cal M}_{LO}^c
\\
{\rm Fig.3c} &\propto&  -g_s \frac{2k^\mu}{2l \cdot k} e^{-iz_\perp \cdot l_\perp} {\cal M}_{LO}^b
 f^{abc}
\end{eqnarray}
where the leading order amplitude is denoted as ${\cal M}_{LO}^c$ with $c$ being the color index of the
gluon $k$. For  the conjugate diagrams, one has,
\begin{eqnarray}
&&ig_s \frac{2 n^\mu}{2l \cdot n }  e^{iy_\perp \cdot l_\perp}t^a{\cal M}_{LO}^{*c} ,
\\
&&-i g_s  \frac{2n^\mu}{2l \cdot n} e^{iy_\perp \cdot l_\perp} {\cal M}_{LO}^{*c} t^a,
 \\
 && -g_s \frac{2k^\mu}{2l \cdot k} e^{iz_\perp \cdot l_\perp} {\cal M}_{LO}^{*b}f^{abc}
\end{eqnarray}
The virtual graphs with a soft gluon emission/absorbtion
 can be readily computed in the same approximation as well.
Using these derived amplitudes for soft gluon radiations
 as the basic calculation ingredients,
we are ready to extract the double leading logarithm contributions from the NLO correction to the BK kernel.

\subsection{The complete and incomplete cancellations between real graphs and virtual graphs }

\begin{figure}[htpb]
\includegraphics[angle=0,scale=0.55]{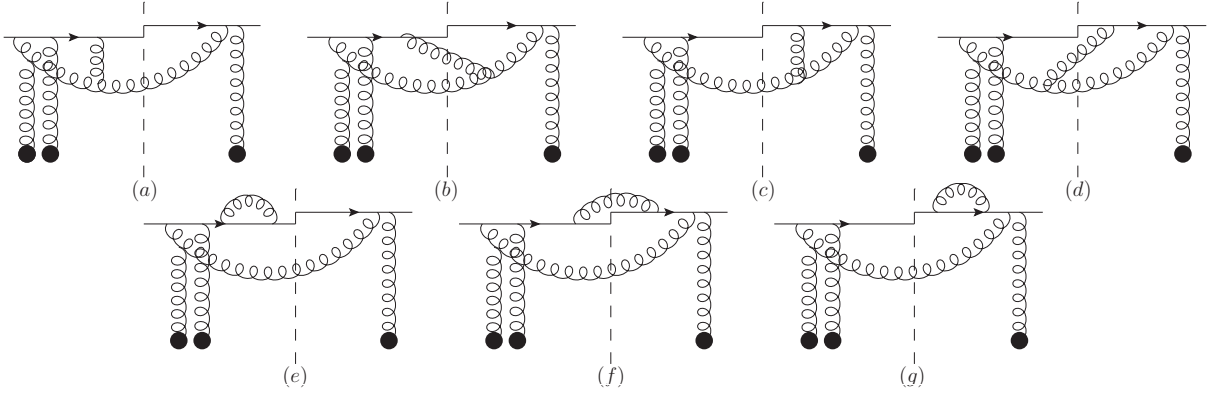}
\caption{  Sample diagrams contributing to the BK kernel at two loop order.
 The contributions from the Figs.[a+b] are completely canceled out up to the double
 leading logarithm accuracy.
 The same cancellation occurs between the  Fig.[c]  and the Fig.[d].
  In contrast, the double leading logarithm NLO correction to the
  real part of the LO BK kernel arise from the incomplete cancellation among
  Fig.e, Fig.f and Fig.g.  } \label{realcan}
\end{figure}

It is well known that the Sudakov type double logarithm results from the mismatch between
soft gluon radiation contributions in real corrections and virtual corrections.
In this subsection, we are going to demonstrate that
 at NLO, such incomplete cancellation indeed occurs in the phase space region  $ k^+ \gg l^+$ and $ k^-\gg l^-$
 for the diagrams with a soft gluon attaching to the Wilson line.
 In contrast, if the soft gluon($l$) is radiated from the hard gluon($k$) line, there is a complete
 cancellation up to the accuracy of interest
  between real diagrams and virtual diagrams as illustrated in Figs[\ref{realcan}a-\ref{realcan}d]

We  first use the Fig[\ref{realcan}a] and Fig.[\ref{realcan}b] as an example to show how the leading double logarithm
contributions are canceled out. This is essentially due to the fact that the virtual contribution
Fig[\ref{realcan}a] and real contribution Fig.[\ref{realcan}b]
 share a common phase factor $e^{-il_\perp \cdot(x_\perp-z_\perp)}$.
 We start with the calculation of the Fig[\ref{realcan}a]. We first
carry out $l^-$ integration  with the residue theorem by picking up the
 contribution from the pole $1/(l^2+i\epsilon)$. After having done so, the soft gluon momentum($l$)
 is effectively put on shell.
  Applying the soft gluon approximation, the virtual correction from  Fig.[\ref{realcan}a] reads,
\begin{eqnarray}
{\rm Fig.[\ref{realcan}a]}&\propto& \!\int \! \frac{dk^+}{ k^+}
\!\int \! d^2 k_\perp \! \int^{|k_\perp|}_0
\! d^2 l_\perp
\!\int^{k^+}_{\frac{l_\perp^2}{k_\perp^2}k^+ }
\!  \frac{d l^+}{l^+} \!\int \!\! d^2 k_\perp' d^2 z_\perp \int d^2 p_{g\perp} d^2 p_{g\perp}' d^2 p_{g1\perp}
 \nonumber \\&\times& e^{-i(z_\perp-x_\perp)\cdot k_\perp'} e^{i(z_\perp-y_\perp)\cdot k_\perp}
  e^{i(x_\perp-y_\perp)\cdot l_\perp}
 e^{-ix_\perp \cdot p_{g\perp}+iy_\perp \cdot p_{g\perp}'}
 \frac{2
 \left[( k_\perp+l_\perp) \cdot  (k_\perp'\!-p_{g1\perp}) \right ] }
 {(k_\perp+l_\perp)^2 (k_\perp'-p_{g1\perp})^2 }
\nonumber \\&\times&
  \frac{\left [n\cdot \epsilon_l^*\right ] \left [k \cdot \epsilon_l\right ] }
 {\left [l \cdot n \right ]\left [- k \cdot l \right ]}
 {\rm Tr}\left [  U^\dag(p_{g\perp}')t^a  t^b  U(p_{g\perp}\!\!-p_{g1\perp} )
  t^c   \right ] f^{bad}\tilde U(p_{g1\perp})_{dc}
\end{eqnarray}
where we shift momentum $k\rightarrow k+l$. The upper limit for $l_\perp$ integration is chosen to be $|k_\perp|$.
 This kinematical constraint is justified because the soft gluon approximation
 is valid only if $|l_\perp| \ll |k_\perp|$.
 Moreover, after subtracting the $(\text{LO})^2$ contribution which is supposed to be absorbed into
 non-local BK equation, the upper limit and lower limit for $l^+$ integration is determined accordingly.
 By changing the integration variables $k-p_{g1}\rightarrow k''$ and $k+l \rightarrow k''' $,
 the integration over $p_{g\perp},p_{g\perp}', p_{g1\perp}$ can be carried out straightforwardly. After
 these manipulations, one obtains,
  \begin{eqnarray}
{\rm Fig.[\ref{realcan}a]}\!&\propto& \!\int \! \frac{dk^+}{ k^+}
\!\int \! d^2 k_\perp \int^{|k_\perp|}_0
\! d^2 l_\perp
\!\int^{k^+}_{\frac{l_\perp^2}{k_\perp^2}k^+ }
\! \frac{ d l^+}{l^+} \!\int \!\! d^2 k_\perp' d^2 z_\perp
 e^{-i(z_\perp\!-x_\perp)\cdot k_\perp''} e^{i(z_\perp\!-y_\perp)\cdot k_\perp}
 e^{i(x_\perp\!  -z_\perp) \cdot l_{\perp}}
 \nonumber \\&\times&
 \frac{2\left[ k_\perp \!\cdot k_\perp'' \right]}
 {k_\perp^2 k_\perp''^2 }
 \frac{\left [n\cdot \epsilon_l^*\right ] \left [k \cdot \epsilon_l\right ] }
 {\left [l \cdot n \right ]\left [ -k \cdot l \right ]}
 {\rm Tr}\left [  U^\dag(y_\perp)t^a  t^b U(x_\perp)  t^c
  \right ]f^{bad}\tilde U(z_\perp)_{dc}
\end{eqnarray}
where $k'''$ has been renamed as $k$.

Similarly, the contribution from  Fig[\ref{realcan}b] is given by,
\begin{eqnarray}
{\rm Fig.[\ref{realcan}b]}\!\!&\propto& \!\! \int \! \frac{dk^+}{ k^+}
\!\int \! d^2 k_\perp \! \int^{|k_\perp|}_0
\! d^2 l_\perp
\!\int^{k^+}_{\frac{l_\perp^2}{k_\perp^2}k^+ }
\! \frac{d l^+}{l^+} \!\int \!\! d^2 k_\perp' d^2 z_\perp \int d^2 p_{g\perp} d^2 p_{g\perp}' d^2 p_{g1\perp}
 \nonumber \\&\times& \!\! e^{-i(z_\perp\!-x_\perp)\cdot k_\perp'} e^{i(z_\perp\!-y_\perp)\cdot k_\perp}
  e^{i(x_\perp\!-y_\perp)\cdot l_\perp}
 e^{-ix_\perp \cdot p_{g\perp}\!+iy_\perp \cdot p_{g\perp}'}
 \frac{2\left[ k_\perp \cdot (k_\perp'\!-p_{g1\perp}) \right] }
 {((k_\perp\!+l_\perp)^2\!+2k\!\cdot \!l)(k_\perp'\!-p_{g1\perp})^2 }
\nonumber \\&\times&
  \frac{\left [n\cdot \epsilon_l^*\right ] \left [k \cdot \epsilon_l\right ] }
 {\left [l \cdot n \right ]\left [ k \cdot l \right ]}
  {\rm Tr}\left [  U^\dag(p_{g\perp}')t^a  t^b  U(p_{g\perp}\!\!-p_{g1\perp} )
  t^c   \right ] f^{bad}\tilde U(p_{g1\perp})_{dc}
\end{eqnarray}
One notices that the main contribution to the above integration is from the phase space region where
$k\cdot l \rightarrow 0$. We thus can neglect $k \cdot l $ term in the denominator
$1/((k_\perp\!+l_\perp)^2+2k\cdot l)$. By applying the same trick, namely, changing integration variables
$k+l \rightarrow k'''$ and $k'-p_{g1} \rightarrow k''$,
 we can readily integrate out $p_{g\perp},p_{g\perp}', p_{g1\perp}$,
\begin{eqnarray}
{\rm Fig.[\ref{realcan}b]}\!\!&\propto& \!\! \int \! \frac{dk^+}{ k^+}
\!\int \! d^2 k_\perp \! \int^{|k_\perp|}_0
\! d^2 l_\perp
\!\int^{k^+}_{\frac{l_\perp^2}{k_\perp^2}k^+ }
\! \frac{d l^+}{l^+} \!\int \!\! d^2 k_\perp' d^2 z_\perp
e^{-i(z_\perp\!-x_\perp)\cdot k_\perp''} e^{i(z_\perp\!-y_\perp)\cdot k_\perp}
  e^{i(x_\perp\!-z_\perp)\cdot l_\perp}
 \nonumber \\&\times& \!\!
  \frac{2\left[ k_\perp \cdot k_\perp'' \right] }
 {k_\perp^2\! k_\perp''^2 }
  \frac{\left [n\cdot \epsilon_l^*\right ] \left [k \cdot \epsilon_l\right ] }
 {\left [l \cdot n \right ]\left [ k \cdot l \right ]}
  {\rm Tr}\left [  U^\dag(y_\perp)t^a  t^b  U(x_\perp )
  t^c   \right ] f^{bad}\tilde U(z_\perp)_{dc}
\end{eqnarray}
where $k_\perp'''$ has been renamed as $k_\perp$. Here some terms suppressed by the power
of $l_\perp^2/k_\perp^2$ have been ignored.
 At this step, it is  clear to see that contributions from Fig.[\ref{realcan}a] and Fig.[\ref{realcan}b]
 are canceled out in the soft gluon approximation, and thus do not produce double logarithm.
 The same analysis also can be applied to other real/virtual diagrams with soft gluon attaching to
 hard gluon line, for instance Fig.[\ref{realcan}c] and Fig.[\ref{realcan}d]. In the following, we focus on
 investigating the diagrams with soft gluon emitted from the Wilson lines.

We  proceed to compute  the double leading logarithm contributions from the diagrams
Fig.[\ref{realcan}e], Fig.[\ref{realcan}f] and Fig.[\ref{realcan}g].
The overall extra phase factor associated with real diagram Fig.[\ref{realcan}f] is
$e^{il_\perp \cdot(x_\perp-y_\perp)}$, while no non-trivial  phase factor is yielded from the virtual contributions
Fig.[\ref{realcan}e] and Fig.[\ref{realcan}g]. Summing up contributions from Fig.[\ref{realcan}e-g],
 one obtains,
\begin{eqnarray}
&& \!\!\!\!\!\!\!\!\!\!\!\!\!\!\!\!\!\!\!\!\!\!
{\rm Fig.\ref{realcan}[e+f+g]}=
g_s^4 \int d^2 z_\perp \int \frac{dk^+}{2\pi k^+} \int \frac{d^2 k_\perp}{(2\pi)^2}\frac{d^2 k_\perp'}{(2\pi)^2}
e^{-i(z_\perp-x_\perp)\cdot k_\perp'} e^{i(z_\perp-y_\perp)\cdot k_\perp}
\nonumber \\ && \ \ \ \ \ \ \ \ \ \ \ \  \times
\int^{|k_\perp|}_0 \frac{d^2 l_\perp}{(2\pi)^2} \frac{2C_F  }{l_\perp^2}
\left \{ e^{i(x_\perp-y_\perp) \cdot l_\perp} -1 \right \}
\int^{k^+}_{\frac{l_\perp^2}{k_\perp^2}k^+ } \frac{d l^+}{2\pi l^+ }
\nonumber \\ && \ \ \ \ \ \ \ \ \ \ \ \ \times
 \frac{4(k_\perp \cdot \epsilon_{k\perp})(k_\perp' \cdot \epsilon_{k\perp}^* ) }{k_\perp^2k_\perp'^2}
 {\rm Tr}\left [  U^\dag(y_\perp)t^a  U(x_\perp)  t^b
  \right ]\tilde U(z_\perp)_{ab}
\end{eqnarray}
where once again, we only take into account the NLO corrections
in the phase space region  $ k^+ \gg l^+$ and $ k^-\gg l^-$, beyond which
 all leading logarithms from the NLO corrections can be organized into non-local BK equation.
The next step is to carry out  $l^+$ integration. This subsequently yields a $l_\perp$ integration,
\begin{eqnarray}
\int^{|k_\perp|}_0 \frac{d^2 l_\perp}{(2\pi)^2} \frac{1 }{l_\perp^2} \ln \frac{k_\perp^2}{l_\perp^2}
\left \{ e^{i(x_\perp-y_\perp) \cdot l_\perp} -1 \right \}
\end{eqnarray}
which can produce a large double logarithm $  \ln^2 \left [ k_\perp^2 (x_\perp-y_\perp)^2 \right ]$
 so long as $k_\perp^2 \gg 1/(x_\perp\!-y_\perp)^2$. After
 integrating out $k_\perp$ and $l_\perp$ with the help of the following integration formula,
\begin{eqnarray}
&&\int \frac{d^2 k_\perp}{(2\pi)^2}
 \frac{k_\perp \cdot \epsilon_{k\perp}  }{k_\perp^2} e^{i(z_\perp-y_\perp)\cdot k_\perp}
\int^{|k_\perp|}_0 \frac{d^2 l_\perp}{(2\pi)^2} \frac{1 }{l_\perp^2} \ln \frac{k_\perp^2}{l_\perp^2}
\left \{ e^{i(x_\perp-y_\perp) \cdot l_\perp} -1 \right \}
\nonumber \\
&=& \frac{-1}{(2\pi)^2}\frac{1}{4} \frac{i(z_\perp\!-y_\perp) \cdot \epsilon_k}{(z_\perp\!-y_\perp)^2}
 \ln^2 \frac{(z_\perp\!-y_\perp)^2}{(x_\perp\!-y_\perp)^2}
 \  \   \  \ \ \ \ \ \ \ \ \ \left ( |x_\perp\!-y_\perp|>|z_\perp\!-y_\perp| \right )
\end{eqnarray}
  the desired transverse double logarithm shows up.
It is then straightforward to carry out the rest integrations.
 When dust settles, the expression is  organized into the form,
\begin{eqnarray}
&& \!\!\!\!\!\!\!\!\!\!\!\!\! {\rm Fig.\ref{realcan}[e+f+g]}=
-\frac{C_F}{2\pi^3}\alpha_s^2  \int \frac{dk^+}{k^+}
\int d^2 z_\perp
 \ln^2 \frac{(x_\perp\!-y_\perp)^2}{(z_\perp\!-y_\perp)^2}
    \nonumber \\ && \ \ \ \ \ \ \ \  \ \ \ \ \ \ \times
    \frac{(x_\perp-z_\perp) \cdot (y_\perp-z_\perp) }{(x_\perp-z_\perp)^2(y_\perp-z_\perp)^2} \
 {\rm Tr}\left [  U^\dag(y_\perp)t^a  U(x_\perp)  t^b  \right ]\tilde U(z_\perp)_{ab}
 \label{realsud}
\end{eqnarray}
At this point, it  becomes evident that the obtained  double transverse logarithm
 has the typical dynamical origin of a Sudakov problem, as stated before.

 The calculations of the rest graphs can be carried out in a similar way.
Summing up contributions from all diagrams with the soft gluon being radiated from  the Wilson lines,
one obtains,
\begin{eqnarray}
&&\!\!\!\!\!\!\!\!\!\!\!\!\!\!\!\!\!\!\!\!\!
g_s^2 \int \frac{dk^+}{2\pi k^+} \int d^2 z_\perp \int \frac{d^2 k_\perp d^2k_\perp'}{(2\pi)^4}
g_s^2 \!\int_0^{|k_\perp|} \!\! \frac{d^2 l_\perp}{(2\pi)^2}
\int^{k^+}_{ \frac{l_\perp^2}{k_\perp^2}k^+} \frac{d l^+}{2\pi l^+ }
 \frac{2}{l_\perp^2} \!\left [ e^{-i(x_\perp-y_\perp )\cdot l_\perp}-1 \right ] \!
 \nonumber \\&& \ \ \ \ \ \ \ \ \ \ \ \ \ \ \ \ \ \ \ \   \times
{\rm Tr} \left [ \left ( t^a{\cal M}_{sum}^{*c}\! -{\cal M}_{sum}^{*c}t^a \right ) \!
\left ( {\cal M}_{sum}^{c}t^a \! -t^a{\cal M}_{sum}^{c} \right )\right ]
\end{eqnarray}
where ${\cal M}_{sum}^{c}$ denotes the summation of all leading order amplitudes.
 Here the color structure can be  simplified using the Fierz identity,
\begin{eqnarray}
{\rm Tr} \left [ \left ( t^a{\cal M}_{sum}^{*c}\! -{\cal M}_{sum}^{*c}t^a \right ) \!
\left ( {\cal M}_{sum}^{c}t^a \! -t^a{\cal M}_{sum}^{c} \right )\right ]=
 N_c {\rm Tr} \left [ {\cal M}_{sum}^{*c} {\cal M}_{sum}^{c} \right ]\!
- \!{\rm Tr} \left [ {\cal M}_{sum}^{*c} \right ]{\rm Tr} \left [ {\cal M}_{sum}^{c} \right ]
\nonumber \\
\end{eqnarray}
To proceed further, we  ignore the contribution associated with the second color structure
${\rm Tr} \left [ {\cal M}_{sum}^{*c} \right ]{\rm Tr} \left [ {\cal M}_{sum}^{c} \right ]$ which is
suppressed in the large $N_c$ limit.
The next step is to convert the whole expression in momentum space into
that in coordinate space by integrating out
hard and soft gluon transverse momentum. We eventually end up with,
\begin{eqnarray}
&&\frac{\alpha_s}{2 \pi^2}\int \frac{dk^+}{k^+} \int d^2 z_\perp
\frac{2(x_\perp-z_\perp) \cdot (y_\perp-z_\perp)}{(x_\perp-z_\perp)^2 (y_\perp-z_\perp)^2}
\left  [-\frac{\alpha_s N_c}{4 \pi}
\left (\ln^2 \frac{(x_\perp\!-y_\perp)^2}{(z_\perp\!-y_\perp)^2}
+\ln^2 \frac{(x_\perp\!-y_\perp)^2}{(z_\perp\!-x_\perp)^2} \right ) \right ]
\nonumber \\ && \times
\left \{ N_c {\rm Tr}\left [  U^\dag(y_\perp)  U(x_\perp) \right ]
- {\rm Tr}\left [  U^\dag(y_\perp)U(z_\perp) \right] {\rm Tr}\left [ U^\dag(z_\perp) U(x_\perp) \right ]
\right \}
\label{realsud1}
\end{eqnarray}
In the limit $(x_\perp-y_\perp)^2\sim (y_\perp-z_\perp)^2 \gg (x_\perp-z_\perp)^2$ or
 $(x_\perp-y_\perp)^2\sim (x_\perp-z_\perp)^2 \gg (y_\perp-z_\perp)^2$,
 the above expression can be cast into a more compact form,
 \begin{eqnarray}
&&\frac{\alpha_s}{2 \pi^2}\int \frac{dk^+}{k^+} \int d^2 z_\perp
\frac{2(x_\perp-z_\perp)\! \cdot \! (y_\perp-z_\perp)}{(x_\perp-z_\perp)^2 (y_\perp-z_\perp)^2}
\left  [-\frac{\alpha_s N_c}{4\pi}
\ln^2 \frac{(z_\perp\!-x_\perp)^2}{(z_\perp\!-y_\perp)^2}
 \right ]
\nonumber \\ && \times
\left \{ N_c {\rm Tr}\left [  U^\dag(y_\perp)  U(x_\perp) \right ]
- {\rm Tr}\left [  U^\dag(y_\perp)U(z_\perp) \right] {\rm Tr}\left [ U^\dag(z_\perp) U(x_\perp) \right ]
\right \}
\label{real1}
\end{eqnarray}
which is the NLO correction to the real part of the BK kernel in the double  leading logarithm approximation.
This result again exhibits a typical factorization structure for a Sudakov problem:
  the BK dynamics  is developed via emitting a hard gluon($k$),
 while the effect induced by the soft gluon radiation is factorized into the double logarithm.

\begin{figure}[htpb]
\includegraphics[angle=0,scale=0.54]{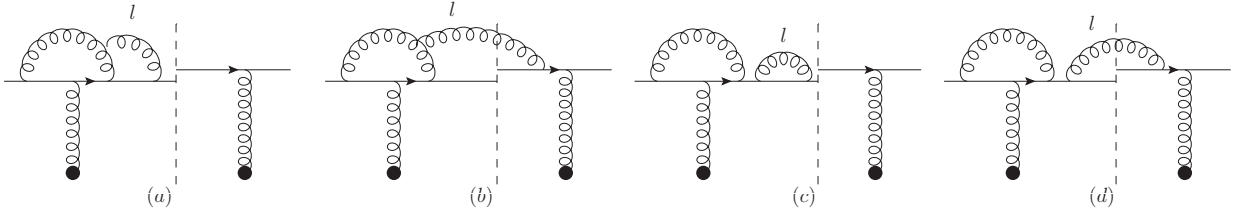}
\caption{ Sample diagrams contributing to the BK kernel at two loop order.
 The different pole contributions from Fig.a and Fig.b  are completely canceled out
  within the double leading logarithm accuracy, while the incomplete cancellation between
  Fig.c and Fig.d gives rise to the partial double leading logarithm NLO correction to the
  virtual part of the LO BK kernel.     } \label{virtualcan}
\end{figure}

Let us now turn to analyze the NLO correction to the virtual part of the leading order BK kernel.
For the virtual graph Fig.[\ref{virtualcan}b], its contribution can be written as,
\begin{eqnarray}
\int
\frac{d^4k}
{[k^2+i\epsilon][(k+l)^2+i\epsilon][k^-+l^--i\epsilon][k^--i\epsilon]
} H(k,l, p_{g\perp}, p_{g\perp}')
\end{eqnarray}
where $k^-$ poles are explicitly shown in the above formula.
$H(k,l, p_{g\perp}, p_{g\perp}')$ represents the rest part of the amplitude.
One can carry out contour integration on $k^-$ by picking up contributions from two poles
 $1/(k^2+i\epsilon)$ and $1/((k+l)^2+i\epsilon)$, respectively,
\begin{eqnarray}
-i2\pi\int d^4k \left [ \delta(k^2)-\delta(k^2+2k\cdot l) \right ]
\frac{H(k,l, p_{g\perp}, p_{g\perp}')}
{[2 k\cdot l][k^-+l^--i\epsilon][k^--i\epsilon]}
\end{eqnarray}
Obviously, two poles contributions cancel out  in the soft gluon limit.
 As a result, this diagram does not produce any  large double logarithm term.
  Such cancellation  occurs for the graph Fig.[\ref{virtualcan}a] as well.
The similar analysis applies to all diagrams with soft gluon attaching to the hard gluon($k$) line.
We thus reach the same conclusion that  the  double leading logarithmic enhanced NLO correction to the virtual part of the LO BK kernel
 can be generated only if soft gluon is radiated from(or absorbed back to) the Wilson lines.
 For instance, the combination of Fig.[\ref{virtualcan}c] and Fig.[\ref{virtualcan}d] generates
 a double logarithm term.
By systematically employing the Eikonal  approximation to all such type diagrams,
it is straightforward to obtain,
  \begin{eqnarray}
&& \!\!\!\!\!\!\!\!\!\!\!\!
-\frac{\alpha_s}{2 \pi^2}\int \frac{dk^+}{k^+} \int d^2 z_\perp
\left \{ \frac{1}{(x_\perp-z_\perp)^2 }
+ \frac{1}{ (y_\perp-z_\perp)^2} \right \}
\left  [-\frac{\alpha_s N_c}{4 \pi}
\ln^2 \frac{(z_\perp\!-x_\perp)^2}{(z_\perp\!-y_\perp)^2}
 \right ]
\nonumber \\ && \ \ \ \ \ \ \ \ \ \ \ \ \ \ \ \ \ \ \ \  \times
\left \{ N_c {\rm Tr}\left [  U^\dag(y_\perp)  U(x_\perp) \right ]
- {\rm Tr}\left [  U^\dag(y_\perp)U(z_\perp) \right] {\rm Tr}\left [ U^\dag(z_\perp) U(x_\perp) \right ]
\right \}
\label{virtualsud}
\end{eqnarray}
where  the  double logarithm term is the same as that in Eq.[\ref{realsud}].
Combining Eq.[\ref{real1}] and Eq.[\ref{virtualsud}] together, one arrives at,
\begin{eqnarray}
 \frac{\partial \langle {\rm Tr} U^\dag(y_\perp)  U(x_\perp) \rangle }{\partial {\rm ln}(1/x_g)}
 \!&=& \!\frac{\alpha_s}{2\pi^2}
  \int\! d^2 z_\perp \frac{(x_\perp-y_\perp)^2}{(x_\perp\!-\!z_\perp)^2(y_\perp\!-\!z_\perp)^2}
  \nonumber \\&\times&\!\!
  \frac{1}{2}\left  [-\frac{\alpha_s N_c}{2 \pi}
\ln^2 \frac{(x_\perp\!-y_\perp)^2}{(z_\perp\!-x_\perp)^2}-\frac{\alpha_s N_c}{2 \pi}
\ln^2 \frac{(x_\perp\!-y_\perp)^2}{(z_\perp\!-y_\perp)^2}
 \right ]
\nonumber \\&\times& \!\!
\left \{
 \langle {\rm Tr}  U^\dag(y_\perp)U(z_\perp) \rangle  \langle {\rm Tr}U^\dag(z_\perp) U(x_\perp) \rangle
- N_c \langle {\rm Tr}  U^\dag(y_\perp)  U(x_\perp) \rangle \right \}
\end{eqnarray}
which is the final double leading logarithmic enhanced NLO correction to the BK kernel.
Note that since our derivation is formulated as  dense target evolution,
 the double leading logarithm term we obtained  is different from that appears
in the full two loop BK equation describing projectile rapidity evolution~\cite{Balitsky:2008zza},
 but instead coincides with the double logarithm contribution derived in Ref.~\cite{Ducloue:2019ezk}
  except for a factor 1/2 difference.
 It is known that the BK equation takes different form at NLO when different
 evolution variables are used.

\begin{figure}[htpb]
\includegraphics[angle=0,scale=0.45]{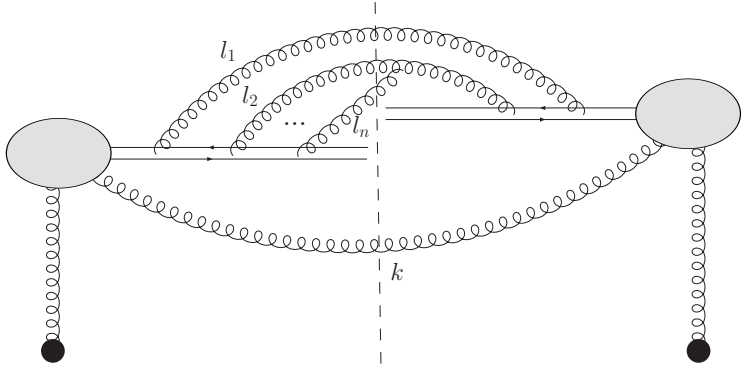}
\caption{ Multiple soft gluon emissions that give rise to the double leading logarithm contributions
 at high order.
 The incoming Wilson line can be effectively treated as the outgoing conjugate  Wilson line.
 A pair of the Wilson line and the conjugate Wilson line in the
fundamental representation  can be viewed as one single Wilson line in the adjoint representation
in the large $N_c$ limit.} \label{final}
\end{figure}

\subsection{Multiple  soft gluon radiations}
When $(x_\perp\!-y_\perp)^2\sim (y_\perp\!-z_\perp)^2 \gg (x_\perp\!-z_\perp)^2$ or
 $(x_\perp\!-y_\perp)^2\sim (x_\perp\!-z_\perp)^2 \gg (y_\perp\!-z_\perp)^2$,
 the large double logarithm compensates the smallness of the
 strong coupling constant $\alpha_s$ and needs to be summed to all orders to improve
 the convergence of the perturbative calculation. To this end, one has to take into account multiple soft gluon emissions
 from the Wilson lines. A schematic graph is shown in Fig[\ref{final}].  To simplify our analysis, we take the
 large $N_c$ limit in which the soft gluon can be viewed as being emitted from a outgoing Wilson line
 in the adjoint representation. To generate a large logarithm like
 $\alpha_s^n \left ( \ln \frac{k_\perp^2(x_\perp-y_\perp)^2}{c_0^2} \right )^{2n}$ at the n-th order, all
 soft gluon momenta $l_1$, $l_2$... $l_n$ must be much smaller than hard gluon momentum $k$.
 However, one does not have to impose a strong ordering in neither soft gluon transverse momenta nor their longitudinal momenta.

All order resummation of the double leading logarithm terms has been achieved in studying the
asymptotic fermion  form factor in QED by Sukakov~\cite{Sudakov:1954sw} more than half century ago.
 The algorithm to resum the Sukakov double logarithm in QCD was later developed in Refs.~\cite{Cornwall:1975ty,Belokurov:1980ba,Sen:1981sd}.
 Such extension   to the non-Abelian gauge theory is highly non-trivial since both
 ladder diagrams and non-ladder diagrams have to be taken into account. In this work, we
  resum these large logarithms by following the Collins' factorization approach~\cite{collins}, based on which
  the soft gluon contribution can be summarized into a
  form factor $S\left (k_\perp^2,(x_\perp-y_\perp)^2 \right )$ whose operator definition is
  the same as that for the soft factor often used in the context of TMD factorization.
  As the hard momentum cut off---$k_\perp^2$ plays the role of $\zeta_c^2$ which is a parameter
  for regularizing the light cone divergence,
 in the spirit of the factorization~\cite{collins},
  we can write down an evolution equation for the form factor in terms of momentum cutoff scale $\mu^2$,
\begin{eqnarray}
\frac{\partial \ln S\left (\mu^2,(x_\perp-y_\perp)^2 \right )}{\partial \ln \mu^2}=
 -\frac{\alpha_s N_c}{ \pi}
\ln \frac{\mu^2 (x_\perp-y_\perp)^2}{c_0^2}
\end{eqnarray}
where $c_0$ is defined as $c_0=2 e^{-\gamma_E}$.
By setting $S\left (1/(x_\perp-y_\perp)^2,(x_\perp-y_\perp)^2 \right )=1$,
 it  can be readily deduced from the evolution equation that,
\begin{eqnarray}
S\left (k_\perp^2,(x_\perp-y_\perp)^2 \right )= \exp \left [ -\frac{\alpha_s N_c}{2 \pi}
\ln^2 \frac{k_\perp^2 (x_\perp-y_\perp)^2}{c_0^2}  \right ]
\end{eqnarray}
where some single logarithm terms are ignored.  With the help of  the integration formula,
\begin{eqnarray}
\int \! \!d^2 k_\perp \!\frac{k_\perp \! \cdot \! \epsilon_k^*}{k_\perp^2} e^{-ik_\perp \cdot (x_\perp\!-z_\perp)}
\!\left [ \ln \frac{k_\perp^2 (x_\perp\!-y_\perp)^2}{c_0^2} \right ]^n \!
= \! -i2\pi \frac{(x_\perp\!-z_\perp)\! \cdot \! \epsilon_k^*}{(x_\perp\!-z_\perp)^2}\!
\left [ \ln \frac{(x_\perp\!-y_\perp)^2 }{(x_\perp\!-z_\perp)^2} \right ]^n\!+...
\end{eqnarray}
 we are able to convert
the exponential into an expression in the coordinate space,
\begin{eqnarray}
\exp \left [ -\frac{\alpha_s N_c}{2 \pi}
\ln^2 \frac{k_\perp^2 (x_\perp-y_\perp)^2}{c_0^2}  \right ]  \longrightarrow
\exp \left [ -\frac{\alpha_s N_c}{2 \pi} \ln^2 \frac{(x_\perp\!-y_\perp)^2}{(z_\perp\!-x_\perp)^2}  \right ]
\end{eqnarray}
or
\begin{eqnarray}
\exp \left [ -\frac{\alpha_s N_c}{2 \pi}
\ln^2 \frac{k_\perp^2 (x_\perp-y_\perp)^2}{c_0^2}  \right ]  \longrightarrow
\exp \left [ -\frac{\alpha_s N_c}{2 \pi} \ln^2 \frac{(x_\perp\!-y_\perp)^2} {(z_\perp\!-y_\perp)^2} \right ]
\end{eqnarray}
Eventually, the Sudakov resummed BK equation takes the form,
\begin{eqnarray}
 \frac{\partial \langle {\rm Tr} U^\dag(y_\perp)  U(x_\perp) \rangle }{\partial {\rm ln}(1/x_g)}
 \!&=& \!\frac{\alpha_s}{2\pi^2}
 \! \int\! d^2 z_\perp \frac{(x_\perp-y_\perp)^2}{(x_\perp\!-\!z_\perp)^2(y_\perp\!-\!z_\perp)^2}
\nonumber \\&\times&
  \frac{1}{2}\! \left \{ \exp\! \left [ -\frac{\alpha_s N_c}{2 \pi} \ln^2 \frac{(x_\perp\!-y_\perp)^2} {(z_\perp\!-y_\perp)^2} \right ]+
  \exp \!\left [ -\frac{\alpha_s N_c}{2 \pi} \ln^2 \frac{(x_\perp\!-y_\perp)^2} {(z_\perp\!-x_\perp)^2} \right ] \right \}
\nonumber \\&\times&
\left \{
 \langle {\rm Tr}  U^\dag(y_\perp)U(z_\perp) \rangle  \langle {\rm Tr}U^\dag(z_\perp) U(x_\perp) \rangle
- N_c \langle {\rm Tr}  U^\dag(y_\perp)  U(x_\perp) \rangle \right \}
\end{eqnarray}
which is our central result of this work. As mentioned earlier, the expansion of this resummed result to the
first non-trivial order  matches to the NLO BK kernel for the double transverse
logarithm term~\cite{Ducloue:2019ezk} apart from an additional factor 1/2 difference.
 Note that the kinematical constraint is not yet explicitly implemented in the above equation.
 One may expect that a Sudakov suppressed non-local BK equation
 effectively resums all important sub-leading logarithms to all orders.
  However, in contrast, it was claimed in Ref.~\cite{Ducloue:2019ezk} that
  such double transverse logarithm is already encoded in the non-local leading order BK equation.
 Obviously, this is not consistent with what we found in the present work.
 But for the time being, we have no clue how
 to solve this mismatch problem, which undoubtedly  deserves further investigation.

\

\section{Summary}
In this paper, we argue that the large transverse double logarithm arises from the NLO correction
to the  BK kernel are the typical Sudakov double logarithm which can be resummed to all orders
following the standard procedure.
 To this end, we first identify the main phase region where the double logarithm is generated.
As recognized recently, to consistently resum small $x$ logarithm $\ln \frac{1}{x_g}$ to all orders,
 the successive gluon emissions must be simultaneously ordered in longitudinal momenta and in lifetimes;
\begin{eqnarray}
k^+ \gg l^+ \gg... \ , \ \ \ \ \ \ \ k^- \ll l^- \ll ...
\end{eqnarray}
   The implementation of  such kinematical constraint results in a  non-local BK evolution equation, which
  effectively resums some sub-leading logarithm terms arise from high order calculations.
  In this work, we explored
  the dynamics beyond this strong ordering region, and found that the large double logarithm can be generated
  in the phase space region where all four components of $l$ is much smaller than that of $k$.
  It is further shown that the resummation of these Sudakov double logarithm can be converted to
  studying the evolution  of  soft factor, which has been well formulated in the literatures~\cite{collins}.
  Therefore, to resum all important sub-leading logarithms, one has to
  include a Sukakov factor in the BK kernel in addition to  imposing a kinematical constraint.
  One may expect that such a resummed BK kernel
  further slows down the small $x$ evolution of dipole.
   We plan to carry out the detailed numerical study of the Sudakov suppressed
  non-local BK equation and investigate its phenomenology implications in a future publication.

\

\

\noindent {\it \bf Acknowledgments:} J. Zhou thanks Feng Yuan for helpful discussions.
This work has been supported by the National Science Foundation of China under Grant No. 11675093,
and by the Thousand Talents Plan for Young Professionals.

\end {document}